%Paper: cond-mat/9310060
%From: Makoto Kaburagi <kabu@icluna.kobe-u.ac.jp>
%Date: Wed, 27 Oct 93 00:31:36 JST

%
%1993-10-25
%ISSP.tex
%

\magnification=\magstep2

\hsize=6.45truein
\vsize=8.89truein
\hoffset=-0.08truein
%\hoffset=-0.36truein
\voffset=0.01truein

\pageno 1
%\nopagenumbers

\baselineskip=16.7pt

\def\vecS{{\vec S}}
\def\vecs{{\vec s}}
\def\lsim{\,$\raise0.3ex\hbox{$<$}\llap{\lower0.8ex\hbox{$\sim$}}$\,}

%\noindent
%RUNNING TITLE: The spin-1 Haldane system with an impurity

\vskip 16pt
\noindent

\centerline{\bf ENERGY VERSUS MAGNETIC-FIELD DIAGRAM OF}

\centerline{\bf THE SPIN-1 HALDANE SYSTEM WITH AN IMPURITY}

\vskip 12pt

\centerline{M. KABURAGI and T. TONEGAWA$^*$}

\vskip 10pt

\noindent
\centerline{Faculty of Cross-Cultural Studies, Kobe University, }
\centerline{Tsurukabuto, Nada, Kobe 657, Japan}

%\vskip 8pt

\noindent
\centerline{$^*$Department of Physics, Kobe University, Rokkodai, Kobe 657,
Japan}

\vskip 12pt

\centerline{ABSTRACT}

\parindent=1.5pc
Energy versus magnetic-field diagram of the spin-$1$ Haldane system
with an impurity bond
is studied in terms of spin-1/2 degree of freedom at the sites
neighboring the impurity bond by means of analytical method. We examine the
equivalence between the realistic Hamiltonian and the phenomenological
Hamiltonian which is composed two spin-1/2 spins representing the spin-1/2
degree of freedom. It is proved that when the strength of the impurity bond
is sufficiently weak, the two Hamiltonians are equivalent to each other,
as far as the energies of the low-lying states are concerned.
We determine
the correspondence between the interaction constants in the phenomenological
Hamiltonian and those in the realistic Hamiltonian.

%\vskip 12pt

\noindent
%KEYWORDS:
%\parindent=3pc
%Haldane System, Impurity, Energy versus magnetic-field diagram.

%\vskip 12pt

%\noindent
%CORRESPONDENCE:
%\parindent=3pc
%Makoto Kaburagi
%
%\parindent=3pc
%\noindent
%(Faculty of Cross-Cultural
%Studies, Kobe University, Tsurukabuto, Nada, Kobe 657, Japan)

%\noindent
%\parindent=3pc
%telefax: 81-78-881-0610,~~~
%(078-881-0610 in Japan)
%\parindent=3pc
%telephone: 81-78-881-1212 (ext.6223),
%(078-881-1212 in Japan)
%\parindent=3pc

%\noindent
%e-mail: kabu@icluna.kobe-u.ac.jp
%\vskip 16pt
%
\vfill\eject
%\bye
%
%1993-10-1
%Hifield.TEX
%

%\magnification=\magstep2

%\hsize=6.45truein
%\vsize=8.89truein
%\hoffset=-0.08truein
%\hoffset=-0.43truein
%\voffset=0.01truein

%\pageno 2

%\baselineskip=16.7pt

%\def\vecS{{\vec S}}
%\def\vecs{{\vec s}}
%\def\lsim{\,$\raise0.3ex\hbox{$<$}\llap{\lower0.8ex\hbox{$\sim$}}$\,}

\parindent=1.5pc
The impurity effects on the quantum antiferromagnetic chain
have been subject of a large number of theoretical and experimental studies.
One of the recent exciting topics of this subject is the edge and/or impurity
effects
on the spin-1 antiferromagnetic Heisenberg chain which is the
simplest system realizing Haldane's conjecture[1] of the difference
between integer-spin and half-integer-spin antiferromagnetic Heisenberg chains.
Kennedy[2] found that the open chain has a fourfold degenerate
ground state composed of a singlet
and a triplet which we call the Kennedy triplet, in contrast to a unique
singlet ground state of the periodic chain[3].  The fourfold degeneracy
of the ground state, which was originally found in the so-called AKLT
model[3] with open boundary conditions, is considered to reflect the
hidden $Z_2\!\times\!Z_2$ symmetry in the open chain[4].  The hidden
symmetry is attributed to the spin-1/2 degree of freedoms at edges
of the chain.
\parindent=1.5pc
Recently the present authors and Harada[5] investigated theoretically the
the impurity-bond effect on the ground state and low-lying excited states of
the chain in terms of domain-wall excitations to interpolate between the
open-chain and the periodic-chain cases.  They showed that the impurity bond
brings about a massive triplet mode in the Haldane gap and that the triplet
state comprises three of the four ground states of the open chain.
Miyashita and Yamamoto[6] perform the Monte Carlo analysis of the open chain
to show that the magnetic moments localized around the edges for the Kennedy
triplet decay exponentially with the decay constant which is about 6 in
lattice spacing. On the experimental side, on the other hand,
Kikuchi {\it et al.}[7] investigated the Zn impurity effects on the
magnetization process of the spin-1 Haldane system NiC$_2$O$_4$2DMIz.
Hagiwara {\it et al.}[8] performed the ESR experiment on
Ni(C$_2$H$_8$N$_2$)$_2$NO$_2$(ClO$_4$), abbreviated NENP,
containing a small amount of spin-1/2 Cu$^{2+}$ impurities
and gave experimental evidence for the existence of the
spin-1/2 degrees of freedom at the sites neighboring the
impurity spin.  They analyzed successfully their experimental results by
using the phenomenological Hamiltonian composed of three spin-1/2
spins with the anisotropic exchange interactions. The present
authors very recently proved the equivalence between the phenomenological
Hamiltonian and a more realistic Hamiltonian and gave a
clear interpretation of the anisotropy of the exchange interaction[9].

\parindent=1.5pc
The purpose of this article is to investigate the energy versus magnetic-field
diagrams of the spin-$1$ Haldane system with an impurity bond
in terms of the spin-1/2 degree of freedom by means of analytical method.
We express the Hamiltonian ${\cal H}$ of the chain with an impurity bond under
the magnetic field as sum of ${\cal H}_0$, ${\cal H}'$ and ${\cal H}_{\rm mf}$
as
$$ \eqalignno{
   {\cal H}& ={\cal H}_0 + {\cal H}'+ {\cal H}_{\rm mf} ~,
                & (1{\rm a})  \cr
   {\cal H}_0& = \sum_{\ell=1}^{N-1} h_{\ell,\ell+1}
         + Jd \sum_{\ell=1}^{N}\bigl(S_\ell^z\bigr)^2 ~, & (1{\rm b})  \cr
   {\cal H}'& = \kappa\,h_{N,1}~,                   & (1{\rm c})  \cr
 & h_{\ell,\ell'}
     = J \bigl(S_\ell^x S_{\ell'}^x + S_\ell^y S_{\ell'}^y
     + \lambda S_\ell^z S_{\ell'}^z\bigr)  \qquad (J,~ \lambda>0)~,~~~~~
                                                        & (1{\rm d})  \cr
      {\cal H}_{\rm mf}
      &=-g (H_x S_{\rm total}^x\!+\!H_z S_{\rm total}^z)~,
                                                       & (1{\rm e})   \cr}
$$
where $\vecS_\ell$ ($\ell\!=\!1$, $2$, $\cdots$, $N$) is the spin-1
operator; $S_{\rm total}^z\!\equiv\!\sum_{\ell=1}^N\!S_\ell^z$ and
$S_{\rm total}^x\!\equiv\!\sum_{\ell=1}^N\!S_\ell^x$; $g$ is the $g$-factor;
$\lambda$ is the parameter which represents the anisotropy of the
exchange; $\kappa$ represents the strength of the impurity bond, with
$\kappa\!=\!0$ and $\kappa\!=\!1$ corresponding, respectively, to the open
chain and the periodic chain. Thus, ${\cal H}_0$, ${\cal H}'$ and
${\cal H}_{\rm mf}$ represent, respectively, the Hamiltonian for the
coupling through the host bonds, that for the coupling through the impurity
bond, and that for the Zeeman energy.

\parindent=1.5pc
In ref.$\,$5, we calculated the energies of the low-lying states
of ${\cal H}_0\!+\!{\cal H}'$ by the variational method, assuming that
$|\kappa|\!\ll\!1$, that is, treating ${\cal H}'$ as a small perturbation.
Using
%the same method as in ref.$\,$5.
this method we calculate the energies of the low-lying states
of ${\cal H}$.
In the Haldane region, to which we confine
ourselves hereafter, the
ground state of the unperturbed Hamiltonian ${\cal H}_0$ is fourfold
degenerate[2,3]. Since the fourfold degeneracy reflects the existence of
the spin-1/2 degree of freedom at the neighboring sites of the impurity bond,
we may expect, as in the previous paper[9], that the energies of the
low-lying states of the present system
can be calculated from the phenomenological Hamiltonian given as
$$ {\cal H}^{\rm phe}
      =  {\bar J} \bigl( s_1^x s_2^x
               + s_1^y s_2^y
               + {\bar {\lambda}} s_1^z s_2^z\bigr)
        - H_{\rm x}{\bar g_{\rm x}} \bigl( s_1^x+s_2^x\bigr)
        - H_{\rm z}{\bar g_{\rm z}} \bigl( s_1^z+s_2^z\bigr)
,  \eqno  (2)
$$
where $\vecs_1$ and $\vecs_2$ are the spin operators which represent the
spin-1/2 degrees of freedom at the sites neighboring the impurity;
${\bar J}$ is the effective exchange constant; ${\bar {\lambda}}$ is the
parameter which represents the anisotropy of the effective exchange;
${\bar g_{\rm x}}$ and ${\bar g_{\rm z}}$ are the effective $g$-factors.
By comparing the energy expressions for ${\cal H}^{\rm phe}$ with those for
${\cal H}$, we will show that, when $|\kappa|<<1$,
${\cal H}^{\rm phe}$ is equivalent to ${\cal H}$, as far as the energies of
the low-lying states are concerned.

\parindent=1.5pc
The variational method discussed in ref.$\,$5, according to which
the Haldane region is given by $4\!>\!d\!>\!2\lambda\!-\!4$, leads to the
following fourfold ground-state wave functions $\Phi$ and $\Phi_N^{(\tau)}$
($\tau=+$, $0$, $-$) expressed in the
matrix-product form[10,11].  The function $\Phi$ describes the state with no
domain wall and is given by
$$ \eqalignno{
  \Phi =&\, {\rm Trace}\bigl[\phi_1 \phi_2 \cdots \phi_{N-1} \phi_N \bigr]~,
                                                             & (3)  \cr
  & \phi_{\ell}
     = \cos(\tilde\theta)\,\zeta_\ell\,\sigma_z
         + {\sin(\tilde\theta)\over\sqrt{2}}
            \bigr(\alpha_\ell\,\sigma_+ + \beta_\ell\,\sigma_-\bigl)~,
                                                             & (4)  \cr}
$$
where $\alpha_\ell$, $\zeta_\ell$, and $\beta_\ell$ are the spin states
at the $\ell$-th site, which correspond, respectively, to $S_\ell^z\!=1$, $0$,
and $-1$, and $\sigma_{\pm}\bigl[=\!(\sigma_x\pm i\sigma_y)/\sqrt{2}\,\bigr]$,
$\sigma_x$, $\sigma_y$, and $\sigma_z$ are the Pauli matrices.  The parameter
$\tilde\theta$ is determined from the equation,
$$
   \cos(2\tilde\theta) = {{d-\lambda}\over{4-\lambda}}~.       \eqno (5)
$$
The function $\Phi_N^{(\tau)}$ describes the states with a domain wall
and is expressed in terms of $\phi_\ell$ and the wall operator $w$ as
$$ \Phi_N^{(\tau)}
     = {\rm Trace}\bigl[\,\phi_1 \phi_2 \cdots
                  \phi_{N-1} \phi_N\,w \bigr]~,                \eqno  (6)  $$
where $w\!=\!-\sigma_-$ for $\tau\!=\!+$, $w\!=\!\sigma_z$ for $\tau\!=\!0$,
and $w\!=\!\sigma_+$ for $\tau\!=\!-$.  It is noted that, when $\lambda\!=\!1$
and $d=0$, $\Phi$ represents the singlet state, and $\Phi_N^{(+)}$,
$\Phi_N^{(0)}$, and $\Phi_N^{(-)}$ represent, respectively, the triplet
(Kennedy triplet) states with $S_{\rm total}^z\!=\!1$, $0$, and $-1$.
It should be noted that a Ne\'el-type configuration of magnetic moment near
the edges is accompanied by the state ${\Phi}_N^{(\tau)}$.
The energy expectation value $E_{\rm s}(\kappa)$ ( $E_{\rm t}^{(\tau)}$ ) of
${\cal H}_0\!+\!{\cal H}'$ with respect to the state $\Phi$ ( $\Phi_N^{(\tau)}$
) is given as
$$ \eqalignno{
     E_{\rm s}(\kappa)
%         &= - (N-1+\kappa)J\Bigl\{\sin^{2}2\tilde\theta+
%                          \lambda\sin^{4}\tilde\theta\Bigr\}
%          + NJd\sin^{2}\tilde\theta~,
%                                                           & (7{\rm a})   \cr
         &= - (N-1+\kappa )J\Bigl\{\Bigl({4-d\over 4-\lambda}\Bigr)
                       \Bigl(1+{d\over 4}\Bigr) \Bigr\}
          + NJd{(4-d)\over 2(4-\lambda)}~.~~
%                                                           & (7{\rm b})   \cr
                                                           & (7)   \cr
     E_{\rm t}^{(0)}
        &\equiv E_{\rm s}(\kappa)+\Delta_0(\kappa)
	                                                         \cr
%        &=  E_{\rm s}(\kappa)+
%	2{\kappa}J\sin^{2}2\tilde\theta~,
%                                                           & (8{\rm a})   \cr
	&= E_{\rm s}(\kappa)+
        \kappa J\,\Bigl\{{2(4-d)(4-2\lambda+d)\over (4-\lambda)^2}\Bigr\}~,
%                                                           & (8{\rm b})   \cr
                                                           & (8)   \cr
     E_{\rm t}^{(\pm)}
        &\equiv E_{\rm s}(\kappa)+\Delta_1(\kappa)
	                                                          \cr
%        &=  E_{\rm s}(\kappa)+
%	{\kappa}J\Bigl\{\sin^{2}2\tilde\theta+
%                          2\lambda\sin^{4}\tilde\theta\Bigr\}~,
%                                                            & (9{\rm a})   \cr
       &=  E_{\rm s}(\kappa)+\kappa J \,\bigl\{ {(4-d)(8+2d-\lambda d)\over
       2(4-\lambda)^2} \bigr\}~.
%                                                            & (9{\rm b})
%%\cr}
                                                            & (9)   \cr}
$$

\parindent=1.5pc
We now calculate the energy versus magnetic field diagram under the same
assumption $|\kappa|\!\ll\!1$ as previous.
By using expression of
the matrix elements given by
$$ \eqalignno{
\hfill \langle\Phi|S_{\rm tot}^z|\Phi\rangle
    = \langle{\Phi}_N^{(0)}|S_{\rm tot}^z|{\Phi}_N^{(0)}\rangle
      = 0 ~,~~~~~~~~~~~~~~~~~~~~~~~~~~~\hfill
                                                &  (10{\rm a})  \cr
\hfill \langle{\Phi}_N^{(+)}|S_{\rm tot}^z|{\Phi}_N^{(+)}\rangle
     = -\langle{\Phi}_N^{(-)}|S_{\rm tot}^z|{\Phi}_N^{(-)}\rangle
      = 1~,~~~~~~~~~~~~~~~~\hfill
                                                & (10{\rm b})  \cr
\hfill \langle{\Phi}_N^{(+)}|S_{\rm total}^x|{\Phi}_N^{(0)}\rangle
    = \langle{\Phi}_N^{(-)}|S_{\rm total}^x|{\Phi}_N^{(0)}\rangle
    = G ~~~~~~~~~~~~~~~~~~~\hfill
                                                          \cr
  = \sin 2\tilde\theta/(1 + \cos^2\!\tilde\theta)
  = 2{\sqrt {(4-d)(4-2\lambda+d)} \over (12+d-4\lambda)}
        ~,~~~\hfill
                                                & (10{\rm c}) \cr
\hfill \langle{\Phi}|S_{\rm total}^x|{\Phi}_N^{(+)}\rangle
 = \langle{\Phi}|S_{\rm total}^x|{\Phi}_N^{(0)}\rangle
  = \langle{\Phi}|S_{\rm total}^x|{\Phi}_N^{(-)}\rangle
  = 0~,~~~~~~~~~~~~~~~\hfill
                                                & (10{\rm d}) \cr
\hfill \langle{\Phi}_N^{(+)}|S_{\rm total}^x|{\Phi}_N^{(-)}\rangle
  = 0~,~~~~~~~~~~~~~~~~~~~~~~~~~~~~~~~~~~\hfill
                                                & (10{\rm e}) \cr}
$$
we obtain the secular equation for the energy difference $\Delta E$ (the energy
measured from $E_{\rm s}(\kappa)$ ) as
$$ \eqalignno{
  \Delta E \Bigl[\bigl[\Delta_0(\kappa)-\Delta E\bigr]
    \bigl[\bigl\{\Delta_1(\kappa)-\Delta E\bigr\}^2-(g H_z)^2\bigl]
      \cr
- 2 (G g H_x)^2 \bigl\{\Delta_1(\kappa)-\Delta E\bigr\}\Bigr]
       = 0~.~~
           & & (11) \cr}
$$
The energy difference $\Delta E$ versus magnetic field $H_x$ or $H_z$
diagram within this approximation can be constructed by solving this equation
[12].

\parindent=1.5pc
Let us examine the relation between ${\cal H}^{\rm phe}$ and ${\cal H}$.
The four energy eigenvalues ${\bar E}_{\rm s}$ and
${\bar E}_{\rm t}^{(\tau )}$ ($\tau=0$, ${\pm}$) of ${\cal H}^{\rm phe}$ for
$H_{\rm x}\!=\!0$ and
$H_{\rm z}\!=\!0$ are calculated to be
$$ \eqalignno{
     {\bar E}_{\rm s}~~~~~~~~~~~
         &= - {{\bar J}\over 4}(2+{\bar\lambda})~,
                                                           & (12)   \cr
   {\bar E}_{\rm t}^{(0)}
        \equiv{\bar E}_{\rm s}+{\bar \Delta}_0
         &={\bar E}_{\rm s}+ {\bar J}~,
                                                            & (13)   \cr
   {\bar E}_{\rm t}^{(\pm)}
        \equiv {\bar E}_{\rm s}+{\bar \Delta}_1
        &= {\bar E}_{\rm s}+{{\bar J}\over 2}(1+{\bar \lambda})~,
                                                            & (14)   \cr}
$$
where the subscripts ${\rm s}$, ${\rm t}$ and the superscript ${\tau}$ have
the same meaning as those of eqs.$\,$(7) to (9) but for the two-spin-1/2
system.  The equation for the energy difference $\Delta{\bar E}$ (the energy
measured from ${\bar E}_{\rm s}$ ) is easily obtained as
$$ \eqalignno{
  \Delta {\bar E} \Bigl[\bigl[{\bar \Delta}_0-\Delta E\bigr]
    \bigl[\bigl\{ {\bar \Delta}_1-\Delta {\bar E}\bigr\}^2-
    ({\bar g}_z H_z)^2\bigl]
      \cr
- 2 ({\bar g}_x H_x)^2 \bigl\{ {\bar \Delta}_1-\Delta {\bar E}\bigr\}\Bigr]
       = 0~.~~
           & &(15) \cr}
$$
Comparing eqs.$\,$(12)-(14) and (15) with eqs.$\,$(7)-(9) and (11),
we can determine the correspondence between the interaction constants in
${\cal H}^{\rm phe}$ and those in ${\cal H}$.  The results are
$$ \eqalignno{
  {\bar J} &\leftrightarrow \kappa J \,
  {2(4-d)(4-2\lambda +d) \over (4-\lambda)^2}~,
         & (16) \cr
  {\bar \lambda}
      &\leftrightarrow{\lambda(4-d) \over 2(4-2\lambda+d)}~.
       & (17) \cr
  {\bar g}_{\rm z} &\leftrightarrow g \,
       & (18) \cr
  {\bar g}_{\rm x} &\leftrightarrow  2g
  {\sqrt {2(4-d)(4-2\lambda+d)} \over (12+d-4\lambda)} \,
       & (19)       \cr}
$$
We have thus shown that, ${\cal H}^{\rm phe}$ is equivalent to ${\cal H}$
when $|\kappa|\!\ll\!1$, as far as the energies of the low-lying excited
states are concerned.  Equations~(17)-(19) give a clear explanation for the
origin of the anisotropy in ${\cal H}^{\rm phe}$;
as seen from eqs.$\,$(17)-(19), in the case of
$\lambda\!=\!\lambda'\!=\!1$ the
uniaxial anisotropy $d$ in ${\cal H}$ produces the anisotropy
in ${\cal H}^{\rm phe}$.  This result confirms again the concept of the
spin-1/2 degrees of freedom at edges of the open spin-1 chain.  The use of
${\cal H}^{\rm phe}$ for the semi-quantitative analysis of the experimental
results is also justified.

\parindent=1.5pc
Finally we discuss the solutions of eq.$\,$(11). The equation has a trivial
solution
$$ \eqalignno{
\Delta E\! &= \!\Delta E_s\!=\!0.  & (20) \cr
}
$$
The three solutions other than this trivial solution
, which we denote as
$\Delta E_t^{(\tau)}(\kappa)$ ($\tau = 0, \pm$), are obtained
for the case of $H_x = 0$ and $H_z\!=\!H\!>\!0$ as
$$ \eqalignno{
     \Delta E_t^{(0)}(\kappa) &= \Delta_0(\kappa)~,             & (21)  \cr
     \Delta E_t^{(\tau)}(\kappa) &= \Delta_1(\kappa) + \tau g_{\rm h}H_z~,~
     ~~~~~~(\tau = \pm)~.~                                 & (22)  \cr}
$$
Those for the case of $H_x > 0$ and $H_z = 0$ are obtained as
$$ \eqalignno{
     \Delta E_t^{(0)}(\kappa)
     = \Delta_1(\kappa)~,~~~~~~~~~~~~~~~~~~~~~~~~~~~~~~~~~~~~~~~~~~~~~~~~~
                                            &~& (23)  \cr
   \Delta E_t^{(\tau)}(\kappa)
     ={\Delta_1(\kappa) +\Delta_0(\kappa) \over 2}+
     ~~~~~~~~~~~~~~~~~~~~~~~~~~~~~~~~~~~~~~~~~
                                              &~&        \cr
      \tau \sqrt{\Bigl({{\Delta_1(\kappa) -\Delta_0(\kappa)}
      \over 2}\Bigr)^2 + (G g H_x)^2}~~~~~(\tau=\pm).~
                                                                 &~& (24)  \cr}
$$
In Fig.$\,$1 we show the $\Delta E$ versus magnetic-field diagram
calculated from either eqs.$\,$(21)-(22),  eqs.$\,$(23)-(24), or eq.$\,$(11)
for the case of $d\!>\!0$ and ${\kappa}\!>\!0$. It should be noted that
we can calculate the magnetic-field or temperature dependence of
the Ne\'el-type configuration of magnetic
moment around impurity bond at low temperatures and that the results
obtained in the present paper are in qualitatively good agreement
with those
obtained by the numerical diagonalization. Details of the calculations
will be published in the near future.

\parindent=1.5pc
The present work has been supported in part by a Grant-in-Aid for
Scientific Research on Priority Areas, ^^ ^^ Computational Physics as a New
Frontier in Condensed Matter Research'', from the Ministry of Education,
Science and Culture.  One of the authors (M.~K.) gratefully acknowledges the
support of Fujitsu Limited.

\vfill\eject
\centerline{\bf References}

\item{1]} F.~D.~M.~Haldane:~Phys.~Lett.~{\bf 93A} (1983)
464;~Phys.~Rev.~Lett.~{\bf 50} (1983) 1153.

\item{2]} T.~Kennedy:~J.~Phys.~Condens.~Matter {\bf 2} (1990) 5737.

\item{3]} I.~Affleck, T.~Kennedy, E.~H.~Lieb and
H.~Tasaki:~Phys.~Rev.~Lett.~{\bf 59} (1987) 799;~Commun.~Math.~Phys.~{\bf 115}
(1988) 477.

\item{4]} T.~Kennedy and H.~Tasaki:~Commun.~Math.~Phys.~{\bf 147} (1992) 431.

\item{5]} M.~Kaburagi, T.~Tonegawa  and I.~Harada:~J.~Phys.~Soc.~Jpn.~{\bf 62}
(1993) 1848.

\item{6]} S.~Miyashita and S.~Yamamoto:~Phys.~Rev.~B {\bf 48} (1993) 913.

\item{7]}  H.~Kikuchi, Y.~Ajiro, T.~Goto, H.~Aruga~Katori and H.~Nagasawa:
these proceedings

\item{8]}  M.~Hagiwara, K.~Katsumata, I.~Affleck, B.~I.~Halperin and
J.~P.~Renard:~Phys.~Rev.~Lett.~{\bf 65} (1990) 3181,
\item{} M.~Hagiwara:~Dr.~Thesis, Graduate School of Science, Osaka
University, Toyonaka, Osaka, 1992.

\item{9]} M.~Kaburagi and T.~Tonegawa: Submitted to ~J.~Phys.~Soc.~Jpn.
%~{\bf 62} (1993) 1848.

\item{10]} A.~Kl\"umper, A.~Schadschneider and J.~Zittartz:~J.~Phys.~A
{\bf 24} (1991) L955;~Z.~Phys.~B~{\bf 87} (1992) 281.

\item{11]} M.~Fannes, B.~Nachtergaele and R.~F.~Werner:~Europhys.~Lett.~{\bf
10}
(1989) 633; Commun.~Math.~Phys.~{\bf 144} (1992) 443.

\item{12]} The equations (11) and (12) in ref.$\,$5 should be corrected
as those obtained by putting $g\!=\!1$ in eqs.(11) and (10c).

\vfill\eject

\centerline{\bf Figure Caption}

\vskip 9pt

{\leftskip=1.5pc
\parindent=-1.5pc
Fig.$\,$1.~~Schematic energy difference $\Delta E$ (the energy measured from
$E_{\rm s}$) versus
magnetic-field diagram for the case of $d\!>\!0$ and $\kappa\!>\!0$ obtained
by the present analysis; (a) for $H_x = 0$ and $H_z\!=\!H\!>\!0$,
(b) for $H_x\!=\!H\!>\!$ and $H_z\!=\!0$,
(c) for $H_x\!=\!H_z\!=\!{H \over \sqrt 2}\!>\!0$.
\par}

%\vskip 9pt

\bye